\documentclass[twocolumn,showpacs,preprintnumbers,amsmath,amssymb,letter]{revtex4}

\usepackage{graphicx}
\usepackage{dcolumn}
\usepackage{bm}
\usepackage{subfigure}

\begin{document}

\title{Shocks in Vertically Oscillated Granular Layers}

\author{J. Bougie, Sung Joon Moon, J. B. Swift, and Harry L. Swinney}
 \affiliation{Center for Nonlinear Dynamics and Department of Physics,
{University of Texas, Austin, TX 78712}}

\date{\today}

\begin{abstract}

We study shock formation in vertically oscillated granular layers,
using both molecular dynamics simulations and numerical solutions of
continuum equations to Navier-Stokes order.  A flat layer of grains is
thrown up from an oscillating plate during each oscillation cycle and
collides with the plate later in the cycle.  The collisions produce
layer compaction near the plate and a high temperature shock front
that rapidly propagates upward through the layer.  The shock is highly
time-dependent, propagating through the layer in only a quarter of the
cycle.  We compare numerical solutions of the continuum equations to
molecular dynamics simulations that assume binary, instantaneous
collisions between frictionless, inelastic hard spheres.  The two simulations
yield results for the shock position, shape, and speed that agree
well.  An investigation of the effect of inelasticity shows that the
shock velocity increases continuously with decreasing inelasticity;
the elastic limit is not singular.
\end{abstract}


\maketitle

\section{Introduction}

\subsection{Background}

A successful continuum description of granular media could exploit the
powerful methods of standard fluid dynamics to describe a variety of
granular flow phenomena, but the validity of a continuum
description of granular materials has been questioned and remains an open
problem \cite{kadanoff1999}. Several investigators have proposed equations of
motion for rapid granular flow using continuum fields -- number density $n$,
velocity $\bf{u}\rm$, and granular temperature $T$ (${3\over2} T$ is
the average kinetic energy due to random particle motion)
\cite{haff,campbell,dufty2001}.  In one approach, the interaction
between grains is modeled with binary, inelastic hard sphere collision
operators in kinetic theory to derive granular continuum equations to
Euler \cite{goldshtein1}, Navier-Stokes \cite{jenkinsandrichman}, and
Burnett \cite{selaandgoldhirsch} order.  However, attempts to use such
continuum equations to predict properties of granular media have
mainly been restricted to idealized systems that are difficult to
produce in experiment \cite{haff,goldshtein3, jenkins98, kamenetsky}
and to steady state or asymptotic time limits
\cite{haff,jenkins98, goldshtein3,rericha}.  The viability of these
continuum equations for experimentally realizable, three-dimensional
(3D), time-dependent granular systems has scarcely been examined.  In
this paper, we use 3D inelastic hard sphere
molecular dynamics (MD) simulations as well as 3D simulations of
continuum equations to Navier-Stokes order to investigate the strongly
time-dependent properties of normal shocks in a vertically oscillated
granular layer.

\subsection{Shock waves}

If the Mach number $Ma$ (the ratio of the local mean fluid speed to the
local speed of sound) is greater than unity at the point where a fluid
encounters an obstacle, a compression wavefront is formed near the
object and steepens to form a shock.  In an ideal fluid with no
viscosity or heat conduction, the wavefront steepens until the fields
develop a mathematical discontinuity between the compressed region and
the undisturbed region.  In a viscous, conducting fluid, the fields
vary continuously, but the front steepens until the shock thickness
(the width of the region in which the fields change from the
undisturbed values to the compressed values) is on the order of a mean
free path.

When a supersonic fluid impinges perpendicularly onto a flat,
impenetrable plate, a shock forms normal to the plate.  The normal
component of the mean velocity (with respect to the plate) is
zero at the plate, but the velocity of the undisturbed fluid far from
the plate is still supersonic, flowing towards the plate.  The Mach
number decreases abruptly from greater than unity to near zero as the
plate is approached.  The density, temperature, and pressure also
change rapidly in the region of the shock front.  This normal shock
propagates back upstream in a direction opposite that of the upstream
flow velocity.

This paper concerns normal shocks in vertically oscillating granular
layers.  A distinguishing feature of granular materials is that
granular flows reach supersonic speeds under
common laboratory conditions \cite{haff, rericha, goldshtein2}.  Hence
shock formation, which is achieved only under extreme conditions in
ordinary gases, is commonplace in granular materials. Shock
propagation has been extensively studied in rarefied gases
\cite{courantandfriedrichs,zeldovichandraizer}, but the similarities
and differences between shocks in ordinary fluids and in granular
media is the
subject of ongoing research \cite{rericha, potapovandcampbell,
goldshtein2}. A recent laboratory and molecular dynamics study
examined time-independent behavior of {\it oblique} shocks in granular
flow between two closely
spaced plates, and the results were compared with 2D simulations of
Navier-Stokes order continuum equations \cite{rericha}.

\subsection{Model system}

The system consists of a layer of grains on an impenetrable plate that
oscillates sinusoidally in the direction of gravity with amplitude $A$
and frequency $f$.  The dimensionless peak acceleration is $\Gamma = 4
\pi^2 f^2 A / g$, where $g$ is the acceleration due to gravity.  This
simple oscillating system has been found to yield standing wave
pattern formation \cite{melo}, convection \cite{knight96}, clustering
\cite{falcon99}, steady-state flow fields far from the plate
\cite{brey01}, and shocks~\cite{goldshtein2}.  We examine
shock formation and propagation in layers that are
approximately 9 particles deep as poured except in Sec. III E, where
layer depth is varied.  Throughout this paper, the
phrase ``the layer'' refers to the dense region in which the volume
fraction of particles is greater than $4 \%$ of the random
close-packed volume fraction, $\nu_{max}=0.65$.

When the maximum acceleration of the plate is greater than $g$,
the layer leaves the plate during each cycle.  
When the layer is off the plate, it is cooled by
inelastic collisions between particles, while the particles are
simultaneously accelerated towards the plate by gravity; this leads to
a large mean velocity compared to the sound speed (See Sec. III C).  We show in
Sec. III that when the layer contacts the
plate later in the cycle, the Mach number is much greater than
unity with respect to the plate, i.e., the flow is supersonic.

Shocks
similar to those presented here were found
for a wide range of $\Gamma$ and $f$.  
Here we report results obtained from a particular set of parameter
values: $\Gamma=3$ and
$f=0.095\sqrt{g / \sigma},$ where $\sigma$ is the diameter of a
particle.  This would correspond to a frequency of approximately $30$
Hz for particles with a diameter of $0.1$ mm.  The particles have a
coefficient of restitution $e=0.9,$ except in Sec. III D, where $e$
is varied to investigate the effect of inelasticity on shock
propagation.  The number of particles in the container per unit area
of the bottom plate is $10/\sigma^2$, except in Sec. III E,
where the number of particles is varied to investigate shocks in
layers of different depths.  That is, the layer would be $10\sigma$
deep if the particles were arranged in a simple cubic lattice.  
In actual packings seen experimentally, $10$ particles/$\sigma^2$
corresponds to an average depth of approximately $H=9\sigma$ 
as poured \cite{bizon98}.  For consistency with previous investigations, we
use $H$ (the depth of the layer as poured) to express the depth of the
layer throughout this paper.  

The values of $\Gamma$ and $f$ in this study were examined previously
in experiments and MD studies of spatial pattern formation in shallow
granular layers
\cite{bizon98}.  In the present simulations, pattern formation was intentionally suppressed by considering a 
container smaller than one full wavelength in either horizontal direction
(with periodic boundary conditions at the side walls); the absence of 
patterns permits better visualization of the dependence of the fields
on height $z$.  

We present 3D continuum and MD simulations of the oscillated granular
system, and analyze the behavior of the fields $n,$
$T,$ and $Ma$.  
Section II describes the methods used to simulate and analyze the
oscillatory time-dependent flow.  Section III reports our results, including an examination
of varying inelasticity and layer depth, and Sec. IV presents our conclusions.

\section{Methods}

\subsection{Molecular dynamics simulation}
We use an inelastic hard sphere molecular dynamics simulation
that was validated in previous studies~\cite{bizon98,moon02}
by comparison to experimentally observed standing wave patterns
for varying control parameters.
The collision model assumes instantaneous binary collisions.
Linear momentum is conserved, but energy is dissipated
through inelastic collisions.
As in the continuum model (Sec. II B), the dissipation is assumed to be
characterized by a single parameter, the normal coefficient of restitution
$e$: 
\begin{equation}
e = -v_n^*/v_n,
\end{equation}
where $v_n = ({\bf v}_1 - {\bf v}_2)\cdot
({\bf r}_1 - {\bf r}_2)/|{\bf r}_1 - {\bf r}_2|$, and
${\bf v}$'s and ${\bf r}$'s are velocity and position vectors for
a pair of colliding particles before the collision (no superscript)
and after the collision (with a superscript $*$).
Particles are monodisperse, and $|{\bf r}_1 - {\bf r}_2|$
and $|{\bf r}_1^* - {\bf r}_2^*|$ are identically $\sigma$.
In previous studies, the normal coefficient of restitution
depended on $v_n$ ~\cite{bizon98}.  We investigate the effects of such
a model in Sec. III D.  However, 
to be consistent with the continuum model, we use
a velocity-independent coefficient of restitution for most of
this study.
We also neglect the surface friction between particles as well as
between the particles and the plate.

The particles are constrained between the bottom plate which
oscillates sinusoidally between height $z=0$ and $z=2A$, and a ceiling
which is fixed at $z=200\sigma$.  In the two horizontal directions,
the width is $20\sigma$, and the boundary condition is periodic.
Except in Sec. III E, 3937 particles were used; the number of
particles was chosen to correspond to $H=9\sigma$ and the total mass
of the layer matches with that
in the continuum simulations. In Sec. III E, 1969 particles were used
to correspond to $H=4.5\sigma$, and 5906 particles were used to
correspond to $H=13.5\sigma$.
The container has the same $e$ as the
particles, and the mass of the container is assumed to be infinitely
large compared to that of the granular layer.

Fields are calculated by dividing the container height into small bins
of size $\sigma$.  The number density is the number of particles
found in a bin divided by the volume of the bin.  The number density
$n$ from MD simulations is averaged over the same phase angle of the
oscillatory state for ten cycles of the plate. 
The velocity  $\mathbf{v}$ of each particle is used to calculate the
mean velocity in each bin, $\mathbf{u}=\left<\mathbf{v}\right>$, 
where the brackets
represent the average over all the particles found in a bin at the
same phase angle of the oscillatory state for ten cycles of the plate.
Granular temperature is then defined as 
$T = {1 \over 3} \left< | \mathbf{v}-\mathbf{u} |^2 \right>$.   
Throughout this paper, we use units such that the particles have mass unity.

\subsection{Continuum simulation}

We numerically solve the continuum equations of Navier-Stokes order
proposed by Jenkins and Richman \cite{jenkinsandrichman}.  These
equations were derived for a dense gas composed of frictionless
(smooth), inelastic hard spheres by applying Grad's 13-moment method
to the inelastic Enskog-Boltzmann equation.  This model yields
hydrodynamic equations for number density (or equivalently, volume
fraction $\nu={\pi\over6}n\sigma^3$), momentum, and granular
temperature:

\begin{equation} {\partial n\over\partial t} +
\nabla\cdot(n\mathbf{u})=0, \label{eq:continuity}\end{equation}

\begin{equation}n\left( {\partial\mathbf{u}\over \partial
t}+\mathbf{u}\cdot\nabla\mathbf{u} \right) =
\nabla\cdot \underline{\mathbf{P}}- n
g{\mathbf{\hat{z}}} , \label{eq:momentum}\end{equation}

\begin{equation}{3\over2}n\left({\partial T\over \partial t}+
\mathbf{u}\cdot\nabla T\right) =
-\nabla\cdot \bf{q}\rm+\underline{\bf{P}}:\underline{\bf{E}}-{\gamma},
\label{eq:energy}\end{equation}
where the components of the symmetrized velocity gradient tensor
$\underline{\bf{E}}$ are given by:  $E_{ij}={1\over2}\left({\partial_j
u_i}+{\partial_i u_j}\right).$  The components of the
stress tensor $\bf\underline{P}\rm$ are given by the constitutive
relation: 
\begin{equation} P_{ij}=\left[ -p + (\lambda
-{2\over3}\mu)E_{kk}\right] \delta_{ij}+2\mu
E_{ij}, \end{equation} 
and the heat flux is calculated from Fourier's law:
\begin{equation}\mathbf{q}=-\kappa \nabla T.\end{equation}

To calculate the pressure, we use the equation of state and
radial distribution function at contact proposed by Goldshtein \it et
al. \rm \cite{goldshtein3} to
include both dense gas and inelastic effects: 

\begin{equation} p=n T \left[ 1+2(1+{e}) G(\nu)\right], \label{eq:state}\end{equation}
where

\begin{equation} G(\nu)=\nu g_0(\nu),\end{equation}
and the radial distribution function at contact, $g_0$, is:

\begin{equation}g_0({\nu})=\left[ 1- \left( 
{\nu\over\nu_{max}}\right) ^ {{4\over3}\nu_{max}} \right]^{-1},\end{equation}
where $\nu_{max}=0.65$ is the 3D random close-packed volume
fraction.

Equations ~(\ref{eq:continuity}-\ref{eq:energy}) differ from those
for a compressible, dense gas of elastic particles by the energy loss term 

\begin{equation}
\gamma = {12\over \sqrt{\pi}} (1-e^2) {n T^{3/2}\over\sigma} G(\nu),
\end{equation}
which arises from the inelasticity of collisions between particles.
The bulk viscosity is given by
\begin{equation}\lambda={8\over 3\sqrt{\pi}}n\sigma T^{1/2} G(\nu),\end{equation}
the shear viscosity by
\begin{equation}\mu={\sqrt{\pi}\over6}n\sigma
T^{1/2}\left[{5\over16}{1\over G(\nu)} + 1 + {4\over5}\left(1+{12\over\pi}\right)G(\nu)\right],\label{eq:mu}\end{equation}
and the thermal conductivity by
\begin{equation}\kappa={15\sqrt{\pi}\over16}n\sigma
T^{1/2}\left[{5\over24}{1\over G(\nu)} + 1 + {6\over5}\left(1+{32\over9\pi}\right)G(\nu)\right].\label{eq:kappa}\end{equation}
In marked contrast to incompressible isothermal liquid layers,
$n$ and $T$ for an oscillating granular layer 
vary by orders of magnitude in both space and time
during each cycle of oscillation; hence the viscosities
and thermal conductivity are very far from constant and must be
calculated at each timestep. 

The equations are numerically integrated using a
second order finite difference scheme on a uniform grid in 3D and
first order adaptive time stepping.  
For densities near
close-packed, terms of order ${1/G(\nu)}$ are negligible in
Eqs.~(\ref{eq:mu}) and ~(\ref{eq:kappa}).  
For reasons of numerical stability, these terms were neglected and artificial
dissipation \cite{roache} was introduced in the low-density region above and
below the layer.  
In the bulk
of the layer, where the shock forms and
propagates, the density near the shock is high, and the artificial dissipation
has little effect on the solution.  

As in the MD simulation,
the granular fluid is contained between two impenetrable horizontal
plates at the top and bottom of the container.  We initially conducted the
simulation in a cell with a ceiling that oscillates with the bottom
plate at a height $200\sigma$ above the bottom plate.  For
$H=9\sigma$, where most of these simulations were conducted, an
average of less than one particle per oscillation
reached a height of $z \geq 80 \sigma$ in the MD simulation.  Changing the
height of the cell from $200 \sigma$ to $80 \sigma$ in the continuum
simulation
resulted in a significant speedup in computation time, but did not
change the shock profile.  The fractional root mean square
difference over one cycle between the shock location produced in
the tall cell ($200 \sigma$ high) and the shorter cell ($80 \sigma$
high) was less than $1 \%$.  Therefore, for computational
efficiency, the continuum simulations were
conducted with a ceiling located at a height $80 \sigma$ above the
bottom plate, oscillating with the bottom plate.  In MD simulations of
shallow layers, $H\lesssim5$, many particles reach heights of
$z>80 \sigma$, so we use a cell height of $200 \sigma$ in continuum
simulations for $H\leq 5$ (see Sec. III E).

As in the MD simulations, we use
periodic boundary conditions in the horizontal directions. 
Initial simulations were conducted with a container $20 \sigma$ wide
in the horizontal directions, as in the MD simulation.  For reasons of
computational efficiency, we reduced the horizontal width of the cell
to $11 \sigma$, which produced no significant changes in the
shock behavior.
The width in the horizontal directions is $11 \sigma$ for continuum
simulations throughout this paper.

The continuum equations
are solved in the reference frame of the container, with the container
oscillations accounted for by a sinusoidal external
forcing term in the momentum equation, Eq.~(\ref{eq:momentum}).  
We use impenetrable boundary conditions at the plates: $u_z = 0$ in
the frame of the plate. Boundary conditions
for horizontal velocities and temperature are set to match those found
in the MD simulation.  For most of the
oscillation, their vertical derivatives at the plates were found to be
negligible in the MD simulation, so for simplicity the boundary
conditions used for the continuum simulation were:
$\partial u_x / \partial z = 
\partial u_y / \partial z =0, $
and 
$\partial T / \partial z = 0,$
although these derivatives are not
identically zero in the MD simulation for the entire oscillation
cycle of
the plate.  

\subsection{Shock tracking and shock width}

With each cycle of the plate, a normal shock wave is formed and
propagates through the layer.  This
shock wave separates a compressed region near the plate from the
undisturbed region which is still falling towards the plate.  
Both the undisturbed
region and the compressed region vary in space and time.  
In Sec. III we find that when the shock
is in the layer, granular temperature
increases with height in the compressed region, drops
quickly in the shock, and then increases again in the undisturbed
region (Fig.~\ref{fig1}).  The shock is identified in simulations as the
region in which the
temperature falls from its maximum at the top of the compressed
region to the local minimum at the bottom of the undisturbed region. The
shock width $d$ is defined as the distance between the locations of this
maximum and this minimum.  For purposes of finding the propagation
speed of the shock, the ``shock location'' is defined as the steepest
region of temperature decrease, which is located near the center of
the shock region (shaded regions in Fig.~\ref{fig1}). 

\section{Results}

\subsection{Shock profile during formation and propagation}

\begin{figure}
\scalebox{.65}{\includegraphics{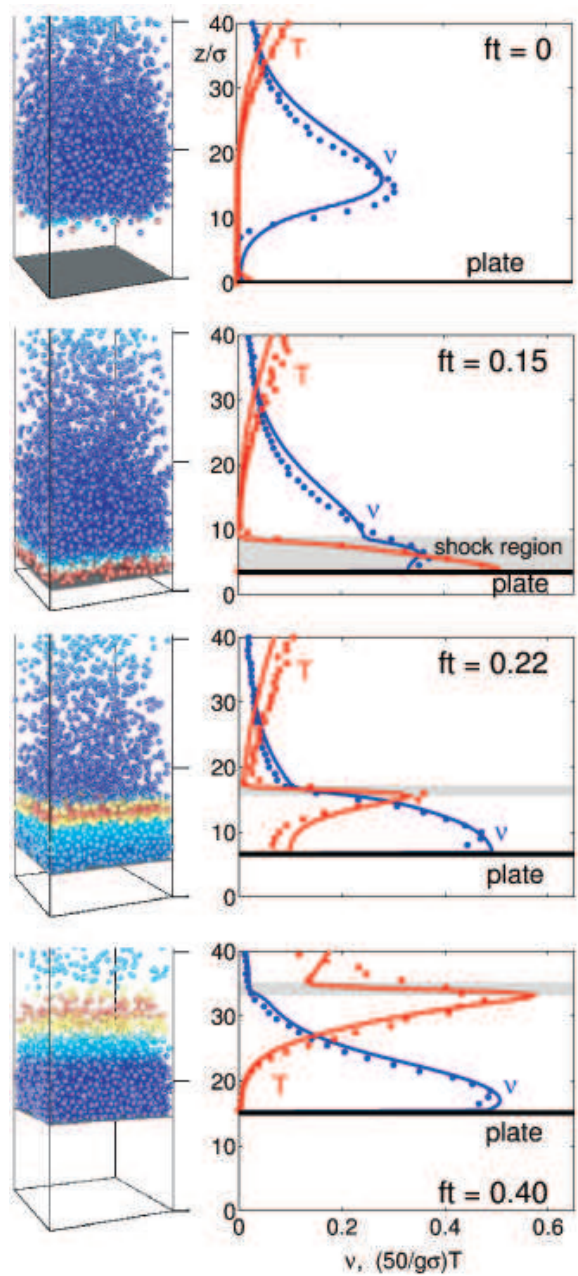}}

\caption{\label{fig1}
Dimensionless temperature $T/(g\sigma)$ and volume fraction $\nu$ 
as functions of dimensionless height $z/\sigma$ at four times $ft$ in
the oscillation cycle.  For each time, a
picture from MD simulation is shown in the left column, with
individual particles color-coded according to temperature: high $T$ in
red, low $T$
in blue, and the bottom plate of the container shaded solid gray.
The right column shows horizontally
averaged $\nu$ (blue, labeled on the bottom graph) and $T/(g\sigma)$
(red, labeled on the top graph) as
functions of $z/\sigma$ (ordinate) for the same four times.  The plate is
shown as a horizontal black solid line, results from MD simulation are
shown as dots, and continuum results are solid lines.  
At times when a shock is present in the
graph, the shock region as located in continuum simulations (see
Sec. II C) is shaded in gray.  The width
of the shock varies throughout the cycle, as
discussed in Sec. III C.}
\end{figure}

Both simulations start with a slightly
randomized flat layer near the plate which is then oscillated.  A
periodic state is reached after a transient
state of several cycles of the plate.  Although most of the particles
move together as a
high-density layer, some particles are always found
above and below the high-density region.  We define ``the
layer'' as the region in which the volume fraction is greater than $4\%$ of
the random close-packed volume fraction $\nu_{max}$.  Below this
volume fraction, there are
too few particles for accurate averaging in the MD simulation.

During each oscillation, the shock forms and propagates at the
same phase angle with respect to the plate's oscillation, so the
dynamics of the
system are fully characterized by the
time interval between time $ft=0$ and one cycle later, $ft=1.$
We now investigate the formation and propagation of the shock by examining
the behaviors found in MD and continuum simulations for
various times $ft$ during the cycle (Fig.~\ref{fig1}).

\subsubsection{$ft=0$: The layer falls towards the plate}

At $ft=0$ the container is at its minimum height.  The particles,
thrown off the plate in the previous cycle, are now falling towards
the plate.  The temperature is nearly zero for most of the material,
increasing with height only in the low density region above the
layer (Fig.~\ref{fig1}).  
The effects of the artificial dissipation which is
used in the continuum simulation are most
pronounced at this time in the cycle, while the layer is dilated and
above the plate. 
The shock
formed in the previous cycle has propagated out through the top of the
layer and is not present in this picture.  The difference between
the MD and continuum simulation at large heights has little effect
on the shock behavior at later times in the cycle.  

\subsubsection{$ft=0.15$: Shock forms as the layer strikes plate}

The layer now begins to strike the rising
plate.  The layer
compresses and particle velocities are randomized by collisions,
increasing $\nu$ and $T$ drastically near the plate.  This
results in the formation of a shock region where $\nu$ and $T$ change
rapidly from the compressed, high temperature
region near the plate to the dilute, low temperature undisturbed region
which is still falling towards the plate.  
At $ft=0.15$ the shock is just beginning to form and is
broad and not fully developed (see Sec. III C).  During this time and
throughout the
formation and propagation of the shock, the shock profile from the
continuum model shows remarkable agreement with the profile found in
the MD simulation.

\subsubsection{$ft=0.22$: Shock propagates through the layer}

The layer continues to compress on the plate, resulting in the
propagation of the shock up through the layer.  Here the shock is
fully developed and has
propagated through much of the layer, but there is still a significant
number of particles falling towards the plate.  The shock has
steepened since $ft=0.15$, with $\nu$ and $T$ changing from the
upstream to the downstream values within a distance of about
$2\sigma.$  As will be discussed in Sec. III C, the width of the shock
continues to change throughout the cycle as the shock propagates
through parts of the layer with different densities.
A discontinuity in
the derivatives of $\nu$ and $T$ appears at the leading edge of the
shock, showing the boundary of the undisturbed region.  Collisions
cause the layer to gradually cool behind the
shock, creating a lower temperature near the plate.

\subsubsection{$ft=0.40$: The layer begins to leave the plate}

The shock has propagated through the bulk of the layer and now enters
the very dilute region above the layer.  
In this region the
temperature becomes uncertain in the MD simulation because
there are too few particles to use in averaging.  Simultaneously, the
plate is approaching
its maximum height, and the layer begins to leave the plate as the
downward acceleration of the plate becomes larger than $g$.  The layer
continues to cool, setting the stage for the next oscillation.

\begin{figure*}
\subfigure{\label{fig2a}\scalebox{.4}{\includegraphics{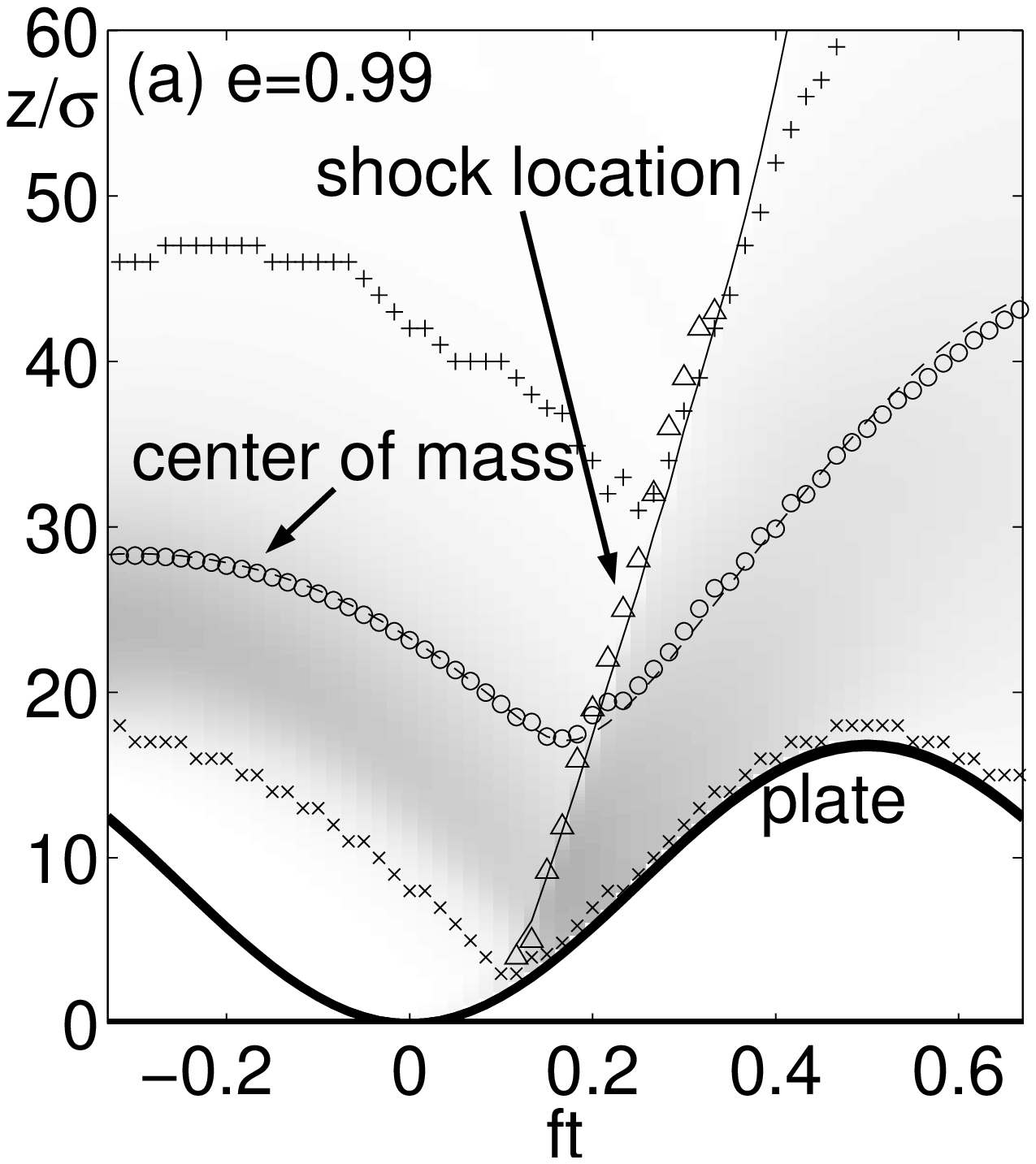}}}
\hspace{-.08in}
\subfigure{\label{fig2b}\scalebox{.4}{\includegraphics{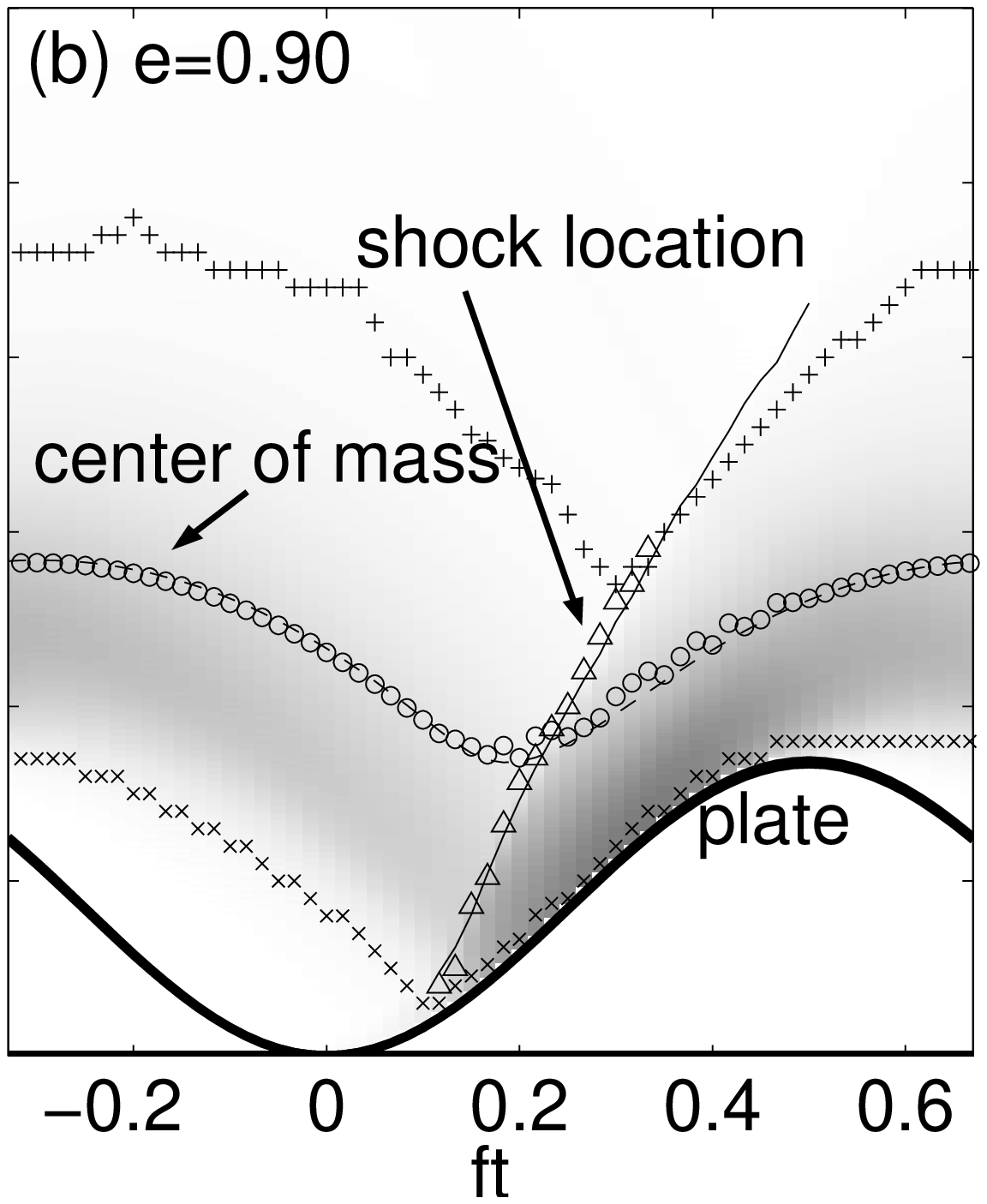}}}
\subfigure{\label{fig2c}\scalebox{.4}{\includegraphics{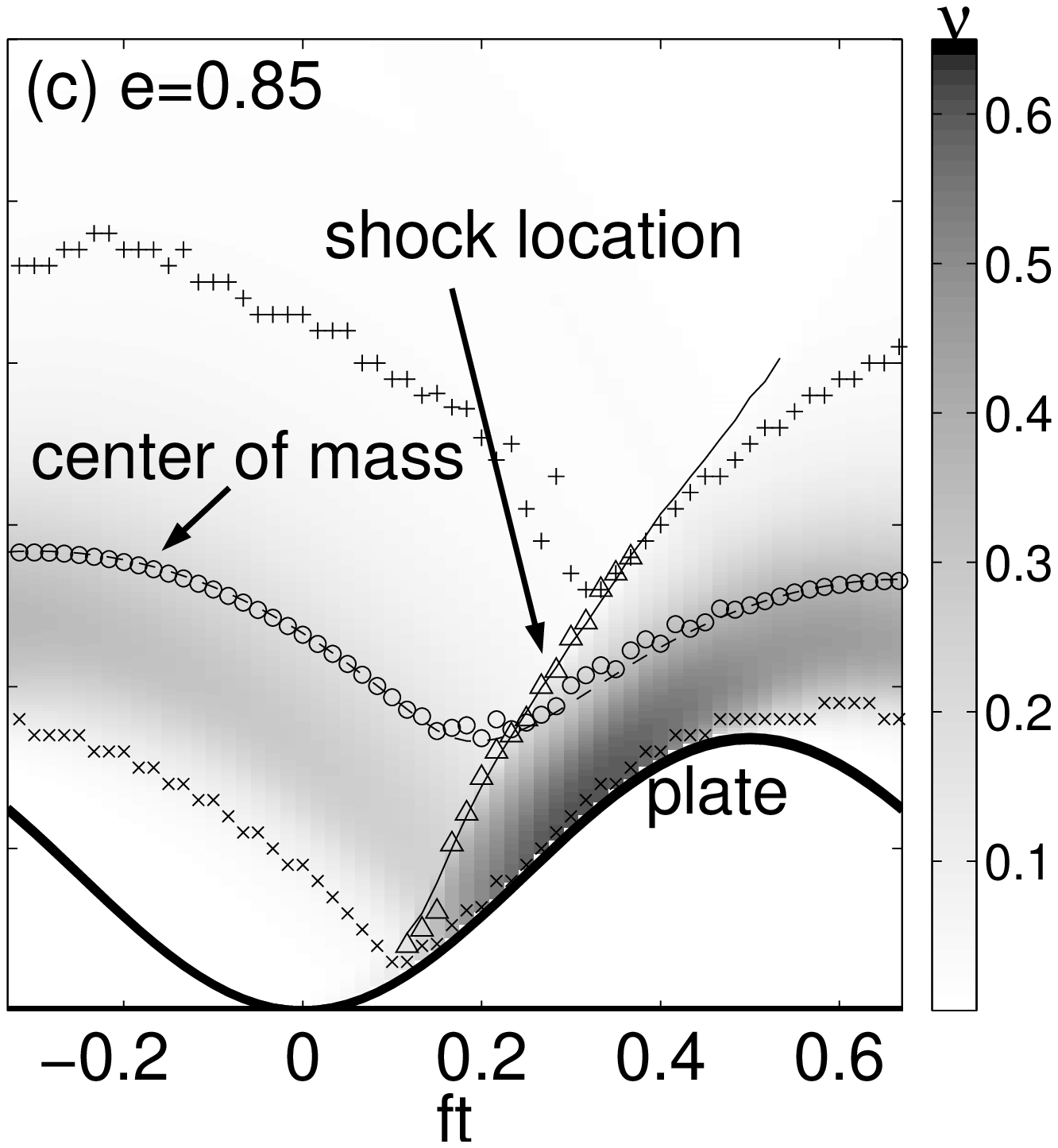}}}

\caption{\label{fig2} Location of the shock (solid lines for continuum,
triangles for MD) and the center of mass of the layer (dashed lines for
continuum, circles for MD) as a function of time $ft$ during one cycle
of the plate (thick solid line) for particles with (a) $e=0.99,$
(b) $e=0.90$, and (c) $e=0.85,$ starting from
the same initial conditions at $ft=-0.33.$
The plot is shaded according to the
volume fraction from the continuum simulation, so that high volume fraction
is dark and low volume fraction is light.  The ``top'' and the ``bottom'' of
the layer from MD (when the volume fraction drops to less than 4\% of random
close-packed) are
shown as +'s and $\times$'s, respectively.  
The material below the
shock is visibly compressed as compared
to the region above the shock, as can be seen from the shading.}
\end{figure*}

\subsection{Shock dynamics}

The formation and propagation of the shock is shown in
Fig.~\ref{fig2} for three values of the restitution coefficient
$e$.  The effect of varying inelasticity will be discussed in Sec. III
D.  In each case, when the bottom of the falling layer hits the
plate, the center of mass of the layer is still falling towards the
plate, and the layer is compressed.  At this time, a
shock is formed near the plate. 
The shock propagates up through the center of mass and out
into the low density region above the layer.  
Then the layer leaves the plate, dilates and cools during its free flight, and
once again falls towards the plate.
As the shock propagates through the layer, it goes through regions of
differing $\nu$ and $T$, and the shock velocity shows
some slight variation with height, as seen in the curvature of
the shock location (Fig.~\ref{fig2}).

\subsection{Mach number and shock width}

A prerequisite for shock formation is that the local Mach number of the
flow be greater than unity with respect to the object causing the
disturbance.  
We calculate the speed
of sound from MD and continuum simulations 
using a relation derived from the continuum equation of state
Eq.~(\ref{eq:state}) \cite{savage}:

\begin{equation}c=\sqrt{T\chi\left(
1+{2\over3}\chi+{\nu\over\chi}{\partial\chi\over\partial\nu}\right)
},\end{equation}
where $\chi = 1+2(1+e)G(\nu).$

\begin{figure}
\scalebox{.4}{\includegraphics{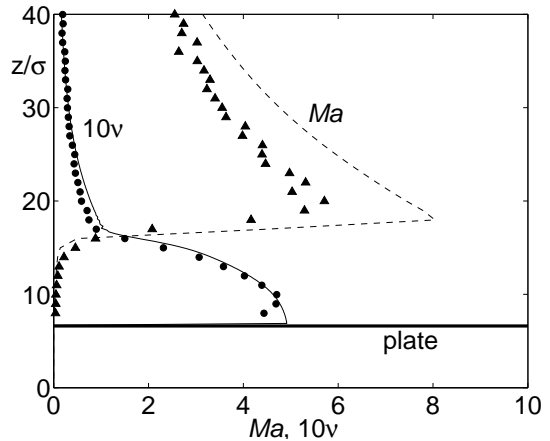}}
\caption{\label{fig3} Mach number ($Ma$) (dashed line for continuum and
triangles for MD) and rescaled volume fraction ($10\nu$) (solid line for
continuum and circles for MD) as
functions of $z/\sigma$ when $ft=0.22$.  At this time the shock is fully
developed and is propagating through the layer (see Fig.~\ref{fig1}).
}
\end{figure}

As the layer falls towards the plate, it is accelerated by gravity
and becomes supersonic in the
reference frame of the plate.  When the layer hits the plate, it is stopped
by the impenetrable boundary, and the Mach number falls to nearly
zero with respect to the plate (Fig.~\ref{fig3}).  The undisturbed region
is still falling towards the plate at supersonic speeds, and the shock
is formed at the boundary between the compressed (low $Ma$) region and
the undisturbed (high $Ma$) region.  The Mach
number in the undisturbed region is largest near the shock where the
layer has been falling the furthest under gravity, and gradually
decreases with increasing height.  
The Mach number is directly calculated from the
values of $\nu,$ $\mathbf{u},$ and $T,$ so the apparent increased
discrepancy between the two simulations as compared to the discrepancy
in previous figures is
due to propagation of errors in these quantities in calculating $Ma.$ 

\begin{figure}
\scalebox{.4}{\includegraphics{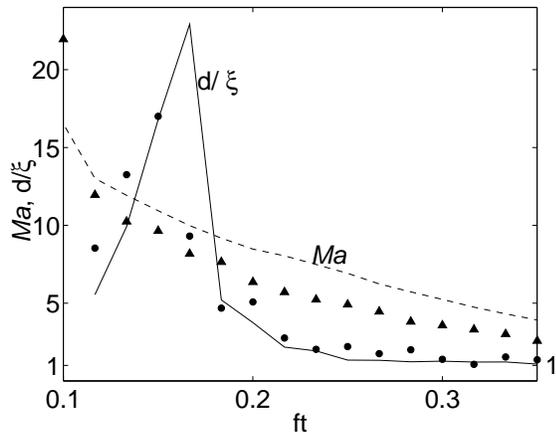}}
\caption{\label{fig4} The maximum Mach number ($Ma$) (dashed line
for continuum and triangles for MD) and the ratio of shock width
$d$ to mean free path
$\xi$ (solid line for continuum and circles for MD) as functions of
time $ft$.  The shock begins forming at $ft=0.12$ and leaves the layer
at $ft=0.35.$}
\end{figure}

While the shock is in the layer, the Mach number always reaches its maximum at
the lower boundary of the undisturbed region.  Thus the maximum value
of $Ma$ at
a given time represents the value at the leading
edge of the shock.  The flows in this system are
hypersonic, with $Ma>10$ just above the plate immediately before the
shock is produced (Fig.~\ref{fig4}).  As the
shock moves through the layer, the maximum $Ma$ decreases since the
fastest flow has been stopped by the shock, but the flow just above
the shock remains supersonic until after the shock leaves the layer
at time $ft=0.35$, and $Ma$ always changes from supersonic to
subsonic as the shock is crossed moving towards the plate.

Since the density of the layer varies in space and time,
the mean free path of the particles in the layer changes as a
function of height and time.  The
mean free path $\xi$ may be estimated as \cite{huang}:

\begin{equation}\xi={\sqrt{\pi/8}\over n \left(\pi \sigma^2
\right)}.\end{equation}
From $ft=0.12$ to $ft=0.17$ the shock is forming but is not fully
developed. The dilute bottom of
the layer hits the plate first at $ft=0.12$, so that while the initial
shock is forming, a higher density region is still falling toward the plate.
As higher and higher density material hits the plate, the
shock width $d$ broadens (Fig.~\ref{fig4}) as it forms
from $ft=0.12$ to $ft=0.17$.  
After $ft=0.17,$ the shock steepens until the fully formed shock
is about one mean free path in width, where it remains until the
shock leaves the layer at $ft=0.35.$  After the shock leaves the layer,
$d$ remains approximately one mean free path, but the
mean free path itself increases in the low density region above the
layer, leading to a broadening of the shock front as the shock leaves
the layer (compare $ft=0.40$ to $ft=0.22$ in Fig.~\ref{fig1}).

\subsection{Effect of inelasticity}

Previous studies have investigated shocks in elastic and inelastic
materials in the asymptotic limit of infinite time \cite{goldshtein3}
and in inelastic materials at finite times \cite{kamenetsky} for a
piston moving into a uniform granular gas at a constant speed in the
absence of gravity.  We wish to investigate the effect of inelasticity on shock
dynamics at finite times in the granular shaker system.  However, the
oscillatory state of the layer is not appropriate for comparing a gas
of elastic hard spheres to one of inelastic hard spheres.
An elastic gas, which has no mechanism for removing energy, will never
reach an oscillatory state since the container injects energy into the
layer at each cycle of the plate.  In order to investigate the effect
of inelasticity on shocks in granular materials, we slightly modify
the system to look at the propagation of a \it single shock \rm from identical
initial conditions rather than looking at oscillatory behavior.

We run simulations with different coefficients of restitution
$e$, starting
from identical initial conditions of a dilated layer falling towards
the plate.  These initial conditions are taken from the oscillatory
state of the system for $e=0.9$ using the same parameters as those
used in previous sections.  At time $ft=-0.33$, when the layer is near
the peak of its flight above the plate, the coefficient
of restitution of the particles is suddenly increased or decreased.  

Changing $e$ significantly
changes properties of the shock and the layer following the next
collision with the plate (Fig. ~\ref{fig2}).  Since the
initial conditions are the same for each case and the layer is in free
fall, very few differences result from the change in $e$ until the
layer collides with the plate.  When the layer hits the plate it is
compacted and many collisions occur between particles.  The lower the
coefficient of restitution, the more energy the particles lose
through these collisions, and the more compact the layer becomes.
In the case of high $e$, the layer rebounds and
dilates quickly after the collision, while in the case of lower
$e$, the layer remains very compact and leaves
the plate almost as a solid body.  The center of mass of the high $e$
layer flies very high before beginning to fall back towards the plate,
while the lower $e$ value leads to a more shallow flight,
with the center of mass beginning to fall earlier
in the cycle.  The shock trajectory exhibits curvature similar to that
of the center of mass, remaining mostly straight for high $e,$ but bending
downward significantly for low $e.$

\begin{figure}
\scalebox{.4}{\includegraphics{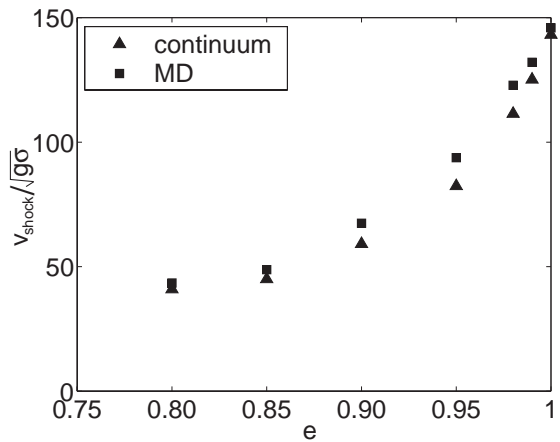}}
\caption{\label{fig5} Average dimensionless shock speed 
$v_{shock}/\sqrt{g\sigma}$ in the reference frame of
the plate as a function of
coefficient of restitution $e$.  $v_{shock}$ is calculated as the average
speed of the shock from when
the shock is formed until it leaves the layer. }
\end{figure}

After the collision with the plate, we track the propagation
of the shock through the layer with a range of new values of $e.$
  Starting from the same initial conditions, the
shock starts at the same time and height regardless of $e.$  However,
the velocity of the shock depends on $e$, 
as Fig.~\ref{fig5} shows.
Although the shock speed changes somewhat as it propagates through the layer
(see Fig.~\ref{fig2}) , the
average shock speed $v_{shock}$ increases
monotonically with $e$ (see Fig.~\ref{fig5}).  The special case of
elastic particles appears to
match with the limit $e \rightarrow 1$; thus there is no qualitative
difference between elastic and inelastic gases.  

\begin{figure}
\subfigure{\scalebox{.4}{\includegraphics{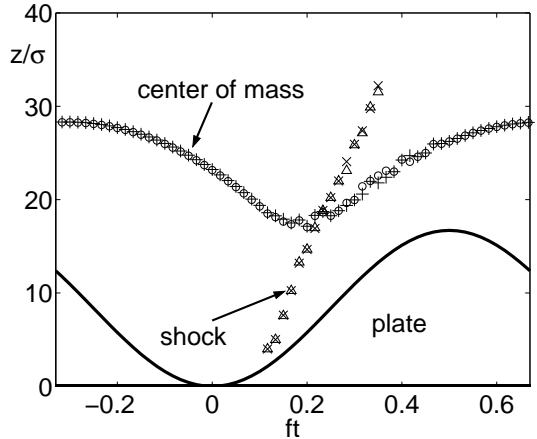}}}
\caption{\label{fig6} 
Location of the shock ($\triangle$ for constant $e=0.9$, $\times$ for
velocity-dependent $e= e\left( v_n \right)$) 
and the center of mass of the layer ($\circ$ for constant $e=0.9$,
+ for velocity-dependent restitution $e= e\left( v_n \right)$) 
as a function of time $ft$ during one cycle
of the plate (thick solid line) in the MD
simulation.
}
\end{figure}

Throughout this paper, both MD and continuum simulations were based on
a collision model in which $e$ was independent of impact velocity. A
more realistic hard-sphere collision model could include a coefficient
of restitution $e=e\left( v_n\right)$ that depends on the
normal component of the relative colliding velocity of the particles, 
$v_n$, such that the
collisions become elastic as $v_n$ approaches
zero.  To test the effect of such a velocity-dependent restituition
coefficient, we ran MD simulations including $e\left( v_n \right)$ of
the same form as that used in studies of
pattern formation in vertically oscillated layers \cite{bizon98}:
$e\left( v_n\right) = 1-0.1\left( v_n/
\sqrt{g\sigma}\right)^{3/4}$ for $v_n < \sqrt{g\sigma}$, and
$e=0.9$ otherwise.  The velocity-dependence of the
coefficient of restitution produces negligible changes in the
shock profiles and trajectories, and less than a 
$1 \% $ change in the average velocity of the shock
(Fig.~\ref{fig6}). For the range of restitution values used in this
paper ($0.8 \leq e \leq 1.0$),
very few particles collide with low enough relative velocity for the
velocity-dependence to significantly affect the outcome.  However, for 
velocity independent $e \lesssim 0.8$, the MD simulation was
unsuccessful due to inelastic collapse, while velocity dependent
$e\left( v_n \right)$ may be used to avoid inelastic collapse with
more inelastic particles.

\begin{figure}
\subfigure{\label{fig7a}\scalebox{.4}{\includegraphics{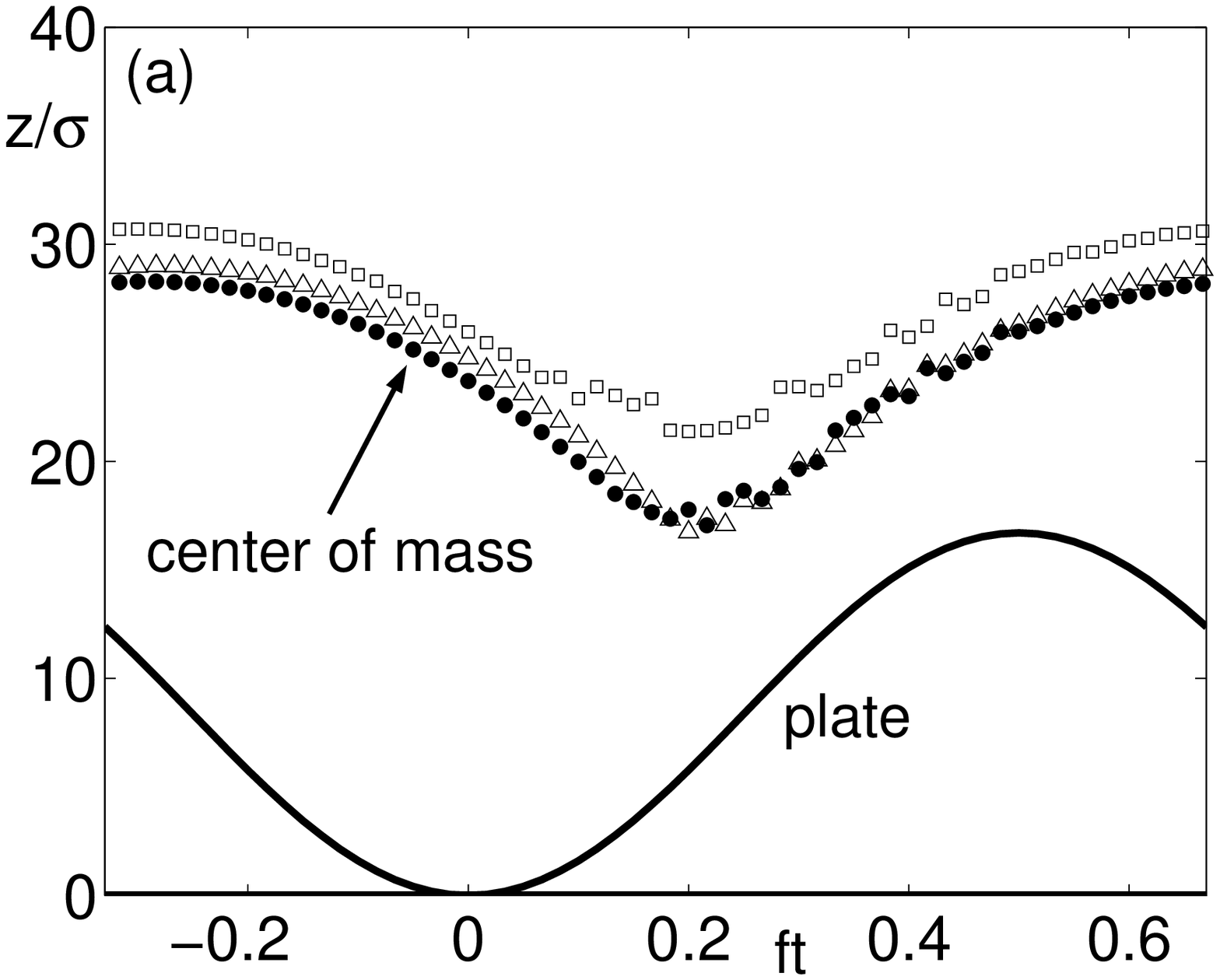}}}\\

\vspace{-.4cm}

\subfigure{\label{fig7b}\scalebox{.4}{\includegraphics{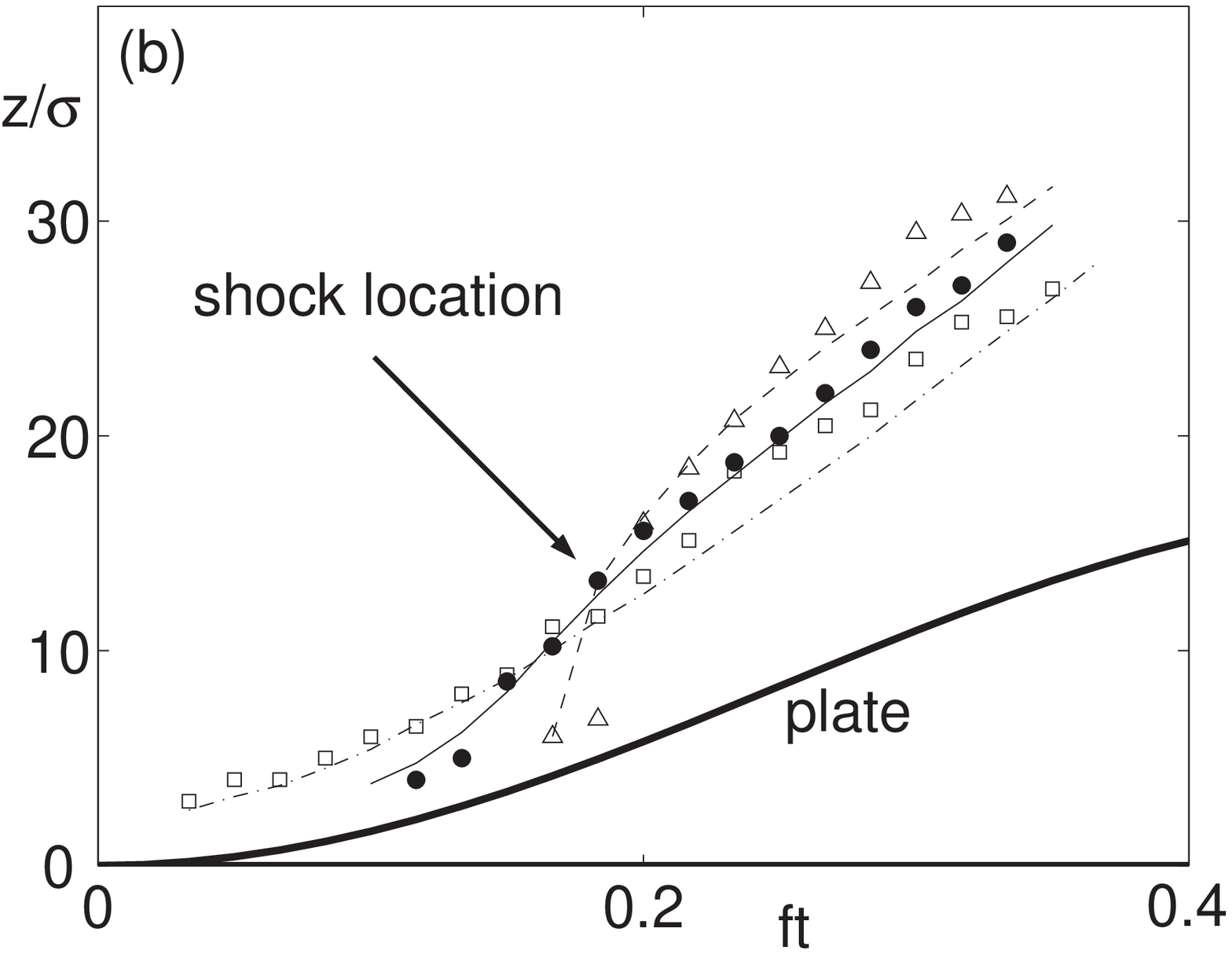}}}
\caption{\label{fig7}
(a) The height of the center of mass of the layer, and (b) the
location of the shock as
a function of time $ft$ for various layer depths $H$. The
trajectory of the center of mass (a) is shown for a complete cycle in
the oscillatory state from the MD simulation with $H=4.5\sigma$
($\square$), $H=9\sigma$ ($\bullet$), and $H=13.5\sigma$
($\triangle$).  The center of mass location from continuum
simulations agrees with MD to within a root mean square difference
of $3\%$ over one cycle in each of the three cases.
The shock location (b) is shown for the fraction of
the cycle in which the shock is in the layer, from the oscillatory
state of $H=4.5\sigma$
($\square$ for MD, $- \cdot -$ for continuum simulation),
$H=9\sigma$ ($\bullet$
for MD, --- for continuum), and $H=13.5\sigma$ ($\triangle$ for MD,
$- - -$ for continuum).}
\end{figure}

\subsection{Shocks in layers of different depths}

Thus far, we have restricted our investigation of shocks in oscillated
layers to layer depths of $H=9 \sigma$.  We
now investigate how changing $H$ affects shock formation and
propagation for particles with $e=0.9$.

For fluidized layers, increasing the layer depth increases the
pressure compressing the
layer on the plate during collision with the plate, resulting in
tighter packing of the particles near the plate.  For $4\sigma \lesssim
H \lesssim 15\sigma$, increasing the layer depth increases the volume fraction
of the layer near the plate, so 
that the trajectory of the center of mass of a layer of depth $H=13.5\sigma$
is nearly identical to that for $H=9\sigma$, even though the former
has 1.5 times as many particles in the layer (Fig.~\ref{fig7a}).  Both
trajectories are actually lower than the center of mass trajectory for
$H=4.5\sigma$.

For $H=4.5\sigma$, the layer is almost entirely gaseous, so much so
that many particles reach heights
greater than $z=80\sigma$ in each oscillation of the plate in MD
simulations with a ceiling at $z=200\sigma$.  Therefore, we use a
cell of height $200 \sigma$ for continuum simulations of $H=4.5\sigma$
(see Sec. II B).  

Changing layer depth also affects the formation and propagation of
shocks in the layer.
For more dilute layers, the shock forms earlier in the cycle, when
the lower edge of the dilute layer collides with the plate.  Thus,
increasing layer depth causes the shock to form later in the cycle, at
which point the shock then propagates faster through the higher
densities found in deeper layers (Fig.~\ref{fig7b}).

For $H=13.5\sigma$,  the pressure on the layer during collision with the plate
is so high that in some small regions near the plate, particles begin to form
ordered lattices and the volume fraction increases
beyond the random close-packed volume fraction.  For deep layers 
$H\gtrsim15\sigma$, a
significant portion of the layer is compressed into a solid
lattice.  The equation of state Eq.~(\ref{eq:state}) does not allow
volume fractions $\nu > \nu_{max}$, so continuum simulations cannot
reproduce densities greater than random close-packed found in
$H\gtrsim15\sigma$.
For $H\lesssim4\sigma$, the particles do not move as a
coherent layer, but rather act as a gas
filling the container for the entire cycle of the plate.

\section{Conclusions}

Shocks form in the vertically oscillated granular system with each
collision of the layer with the plate.  Properties of
these shocks captured in both MD and continuum simulations
show good agreement 
even though the system exhibits a strong time dependence and 
large spatial gradients.
Results from the continuum
simulation agree with the MD results even
for ranges where the validity of the continuum equations is in
question, including when $e$ is
significantly less than one ($e = 0.8$) and when the volume fraction approaches
close-packed.  We also found that deeper layers exhibit denser packing
near the plate and higher shock speeds than shallow layers.  In
addition, adding a simple velocity-dependence to the coefficient of
restitution did not significantly change the behavior of the system.
Finally, we found that increasing the
coefficient of restitution causes 
no singular behavior in the shock velocity
as $e$ approaches unity.

These results demonstrate that continuum equations can describe
even a complicated three-dimensional, time-dependent granular
system.  These results also suggest directions for
further research.  A more realistic model of particle collisions could
include friction between particles.   The simulations should be
further tested by direct comparison to laboratory measurements on
shocks.  Finally, an investigation
of the role of a shock wave in transporting energy from the plate to the layer
could provide
useful information about the mechanism by which oscillations of the
plate provide the energy to drive flows and patterns.

\begin{acknowledgments}
We thank Daniel I. Goldman, W. D. McCormick, and Erin C. Rericha for helpful
discussions.  This work was supported by the Engineering Research Program of the
Office of Basic Energy Sciences of the Department of Energy (Grant
DE-FG03-93ER14312) and a grant from the Texas Advanced Research
Program (Grant ARP 003658-0055-2001).
\end{acknowledgments}

\bibliography{paper}

\end{document}